\documentclass[conference]{IEEEtran}
\usepackage{cite}
\usepackage{amsmath,amssymb,amsfonts}
\usepackage{algorithmic}
\usepackage{graphicx}
\usepackage{tabularx}
\usepackage{float}
\usepackage{subfigure}
\usepackage{xcolor}
\usepackage{stfloats}
\usepackage{makecell}

\usepackage{soul}
\setstcolor{blue}

\begin{document}
\title{Evaluating Power Control Strategies for UORA in IEEE 802.11be Systems with Capture Effect}

\author{\IEEEauthorblockN{
        Kuan-Chin Li\IEEEauthorrefmark{1},
        Ting-Wei Hung\IEEEauthorrefmark{1},
        Lain-Chyr Hwang\IEEEauthorrefmark{2},
        Pengwenlong Gu\IEEEauthorrefmark{3}
        Ray-Guang Cheng\IEEEauthorrefmark{1}
        \\ 
  }
 \IEEEauthorblockA{\IEEEauthorrefmark{1}
Department of Electronic and Computer Engineering, National Taiwan University of Science and Technology, Taiwan
}
\IEEEauthorblockA{\IEEEauthorrefmark{2}
Department of Electrical Engineering, I-Shou University, Taiwan
  }
\IEEEauthorblockA{\IEEEauthorrefmark{3}
CEDRIC, CNAM, France
  }

  Email: crg@mail.ntust.edu.tw
}
\maketitle

%
\begin{abstract}
Uplink OFDMA-based random access (UORA) is a new channel access mechanism that supports uplink multiuser access in the new generation WiFi systems. Any associated stations (STAs) can use UORA to send their requests or data to the access point (AP) in a contention manner. 
In this paper, we provide a comprehensive evaluation for simulation study that investigates two power control strategies combined with capture effect in UORA and observates the fairness issue for spatial distribution of STAs. The results demonstrate that power control strategies can improve the performance of access success probability, delay, resource utilization, and power efficiency of UORA. 
\end{abstract}

\section{Introduction}

IEEE 802.11 networks support multiple channel access mechanisms: Distributed Coordination Function (DCF), Point Coordination Function (PCF), Enhanced Distributed Channel Access (EDCA), and Hybrid Coordination Function Controlled Channel Access (HCCA). 
Contention-based DCF and EDCA dominate Wi-Fi deployments, while contention-free PCF and HCCA are rarely implemented due to complexity and limited flexibility. 
IEEE 802.11ax, with further enhancements in 802.11be (EHT), introduced Triggered Uplink Access (TUA) to improve uplink efficiency via explicit resource allocation by the access point (AP). Within TUA, Uplink OFDMA Random Access (UORA) enables multiple stations (STAs) to contend for resource units (RUs) coordinately, enhancing spectral efficiency and massive device connectivity.

In UORA, the AP allocates random access resource units (RA-RUs) via a trigger frame. STAs select an initial OFDMA backoff (OBO) counter randomly from the OFDMA contention window (OCW) and decrement it by the number of available RA-RUs per trigger frame. When the OBO counter reaches zero, the STA randomly selects one of the RA-RUs and attempts uplink transmission. The AP then acknowledges successful transmissions using a Multi-STA Block ACK. While this mechanism enables efficient uncoordinated uplink access, simultaneous transmissions from multiple STAs can still result in collisions, which degrade performance.

To address these challenges, several studies have proposed enhancements to UORA in IEEE 802.11ax and 802.11be. 
These include cyclic and group resource assignment to reduce latency in large-scale real-time applications~\cite{RTA applications}, adaptive modulation-based $\alpha$-UORA backoff to reduce collisions and improve throughput~\cite{adaptive backoff}, high-performance UORA with a reservation phase for IoT healthcare~\cite{Healthcare IoT}, and double-stage UORA for collision reduction in Wi-Fi 7 IoT networks~\cite{Double stage UORA}. 
Additional schemes comprise centralized AP-controlled $\alpha$ adjustment for efficient OBO adaptation~\cite{Efficient Backoff} and distributed autonomous contention parameter adaptation~\cite{collision based}.
Beyond parameter tuning, Kosek-Szott \textit{et al.} applied deep reinforcement learning to OBO optimization, noting that well-designed heuristics can match or exceed ML approaches~\cite{RL-UORA}. 
Liu \textit{et al.} developed analytical models for saturated/unsaturated UORA, extending to multi-link and multi-RU scenarios~\cite{UORA-optim}. 
Meng \textit{et al.} optimized multi-user aggregation with fixed- and variable-size MPDUs~\cite{MU-UORA}. 
Cheng \textit{et al.} analyzed bursty traffic with queue dynamics~\cite{bursty-11ax}. Choi studied power-domain NOMA with multichannel random access to improve throughput via power diversity and SIC~\cite{NOMA-RA}. 
Szott \textit{et al.} provide a comprehensive survey of ML applications in IEEE 802.11 random access~\cite{WiFi-ML-survey}.
However, prior works have primarily focused on backoff parameter tuning, frame aggregation, or traffic modeling, leaving the interplay between transmit power control and the capture effect in UORA largely unexplored.

In this paper, we present a comprehensive simulation study investigating two power control strategies, random power level selection and stepwise power ramping, combined with the capture effect in IEEE 802.11be UORA networks, with particular emphasis on spatial fairness across concentric rings of equal area using Jain’s Fairness Index. 
Unlike prior works that enhance UORA via backoff optimization~\cite{Efficient Backoff}, reinforcement learning~\cite{RL-UORA}, collision-based distributed control~\cite{collision based}, or analytical backoff modeling~\cite{UORA-optim}, our work focuses on the power domain and analyzes the interaction between power diversity and the capture effect. 
To the best of our knowledge, this is the first study to quantify the fairness implications of capture-aware power control in UORA. 
Simulation results demonstrate that these strategies improve access success probability, delay, resource utilization, and power efficiency.

\section{System Model and Power Control Strategies}
\begin{figure}[!t]
\centering
\includegraphics[width=0.78\linewidth]{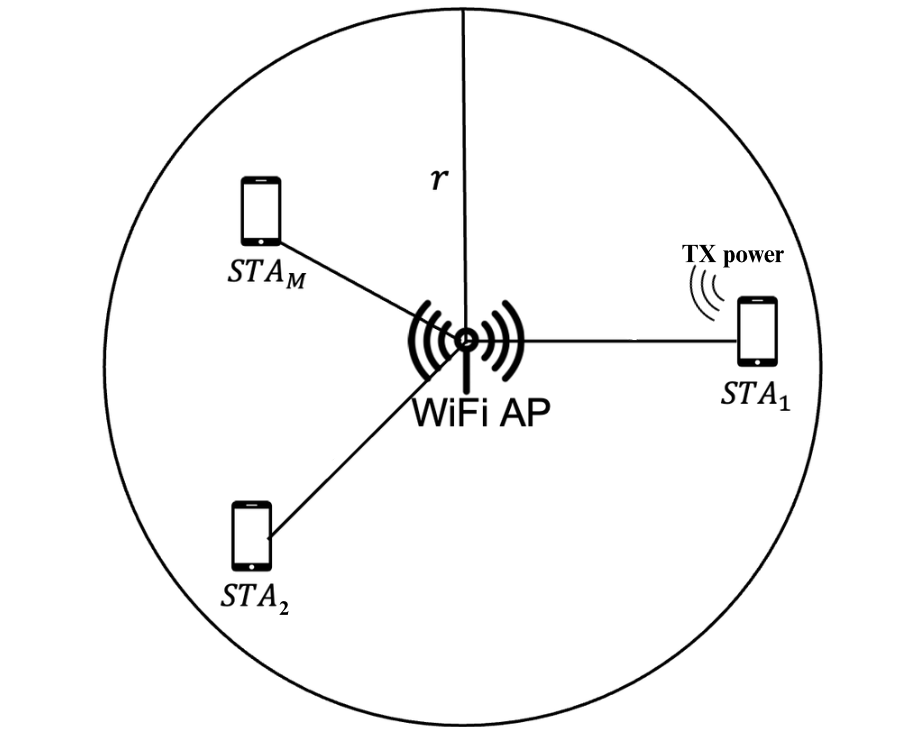}
\caption{Network topology of the considered IEEE 802.11be UORA system. One AP is placed at the center of a circular service area with radius $r$, and $M$ STAs are uniformly distributed in the area. The AP periodically allocates $R$ RA-RUs through trigger frames, while STAs contend for uplink transmission. Due to path loss and transmit-power control, collided STAs can have different received powers at the AP, enabling capture when the strongest SIR exceeds threshold $C$.}
\label{topology}
\end{figure}

We consider an uplink random access scenario in an IEEE 802.11be network consisting of one access point (AP) and $M$ stations (STAs). As shown in Fig.~\ref{topology}, the AP is located at the center of a circular service area with radius $r$, and the STAs are uniformly distributed in this area. At the beginning of the observation interval, each STA has one packet to transmit. The AP periodically broadcasts trigger frames to allocate $R$ random access resource units (RA-RUs), and STAs contend for uplink access through the UORA procedure. Bursty packet arrivals are considered following \cite{bursty-11ax}.

\begin{figure}[!t]
\centering
\includegraphics[width=\linewidth]{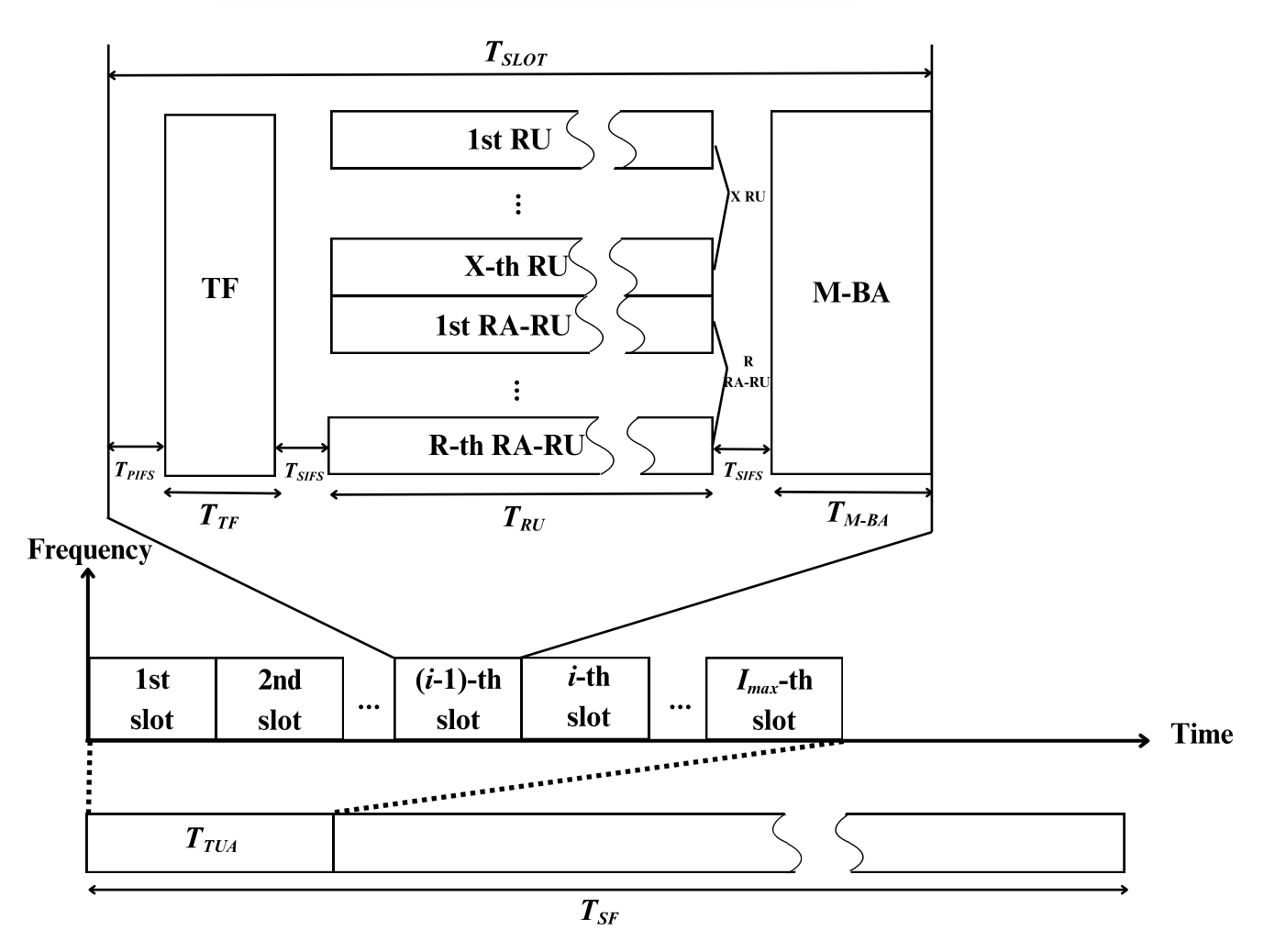}
\caption{Timing structure of the UORA access procedure. Each superframe contains a TUA period followed by EDCA/HCCA operation. During the TUA period, the AP sends a TF to announce $R$ RA-RUs. STAs whose OBO counters reach zero randomly select one RA-RU for uplink transmission, and the AP replies with an M-BA. The slot duration includes PIFS, SIFS, TF, RA-RU transmission, and M-BA intervals.}
\label{system}
\end{figure}

Fig.~\ref{system} summarizes the access timing. Each superframe of duration $T_{\textnormal{SF}}$ includes a triggered uplink access (TUA) period and an EDCA/HCCA period. During TUA, the AP transmits a trigger frame (TF), STAs whose OFDMA backoff (OBO) counters reach zero randomly select one RA-RU, and the AP replies with a multi-STA BlockAck (M-BA). The UORA slot duration is
\begin{equation}
T_{\textnormal{slot}}=T_{\textnormal{PIFS}}+2T_{\textnormal{SIFS}}+T_{\textnormal{TF}}+T_{\textnormal{RA-RU}}+T_{\textnormal{M-BA}} .
\end{equation}
The OFDMA contention window (OCW) used in the $n$-th transmission attempt is
\begin{equation}
OCW_n=\begin{cases}
OCW_{\min}, & n=1,\\
\min(2OCW_{n-1}+1,OCW_{\max}), & n>1.
\end{cases}
\end{equation}

Three transmission schemes are compared. In conventional UORA, all STAs transmit with the minimum power $p_{\min}$. In random power level selection, each STA independently selects its transmit power uniformly from the $L_{\max}$ discrete levels in $[p_{\min},p_{\max}]$ for each transmission attempt. In step-wise power ramping, an STA starts from $p_{\min}$ and increases its transmit power after each failed attempt until $p_{\max}$ is reached. For a collided RA-RU, the AP decodes the strongest STA if its received signal-to-interference ratio (SIR) exceeds the capture threshold $C$.

The performance metrics of access success probability $\mathbb{P}$, average access delay $\mathbb{D}$, channel utilization of RA-RUs $\mathbb{U}$, power efficiency $\mathbb{E}$
 , average power consumption for successful STAs $\mathbb{E}_{\textnormal{S}}$, average power consumption for failed STAs $\mathbb{E}_{\textnormal{F}}$, and average power consumption for all STAs $\mathbb{E}_{\textnormal{T}}$
were investigated during an observation interval of $I_{\textnormal{max}}$ slots.
$\mathbb{P}$ is the ratio of the total number of successful STAs (i.e., STAs that successfully transmit its packet) to the total number of STAs.
$\mathbb{D}$ is the total delay experienced by the successful STAs divided by the total number of successful STAs.
$\mathbb{U}$ is the ratio of the total number of successful RA-RUs to the total number of reserved RA-RUs.
The power efficiency, $\mathbb{E}$, is the ratio of total power consumed by successful STAs to the total power consumed by all STAs.
The access success probability, $\mathbb{P}$, is given by
\begin{equation}
\begin{aligned}
\mathbb{P}  = \frac{\textnormal{Total number of successful STAs}}{\textnormal{Total number of STAs}}.
\label{P}
\end{aligned}
\end{equation}
The average access delay, $\mathbb{D}$, is given by
\begin{equation}
\mathbb{D} = \frac{\textnormal{Total delay experienced by successful STAs}}{\textnormal{Total number of successful STAs}}.
\label{D}
\end{equation}

The channel utilization of RA-RUs, $\mathbb{U}$, is given by
\begin{equation}
\mathbb{U} = \frac{\textnormal{Total number of successful RA-RU}}{\textnormal{Total number of reserved RA-RU}}.
\label{U}
\end{equation}

The power efficiency, $\mathbb{E}$, is given by 
\begin{equation}
    \mathbb{E} = \frac{\textnormal{Total power consumed by successful STAs}}{\textnormal{Total power consumed by all STAs}}. \\
\label{E}
\end{equation}
\begin{figure*}[!t]
\centering
\subfigure[Access success probability]{\includegraphics[width=0.45\textwidth]{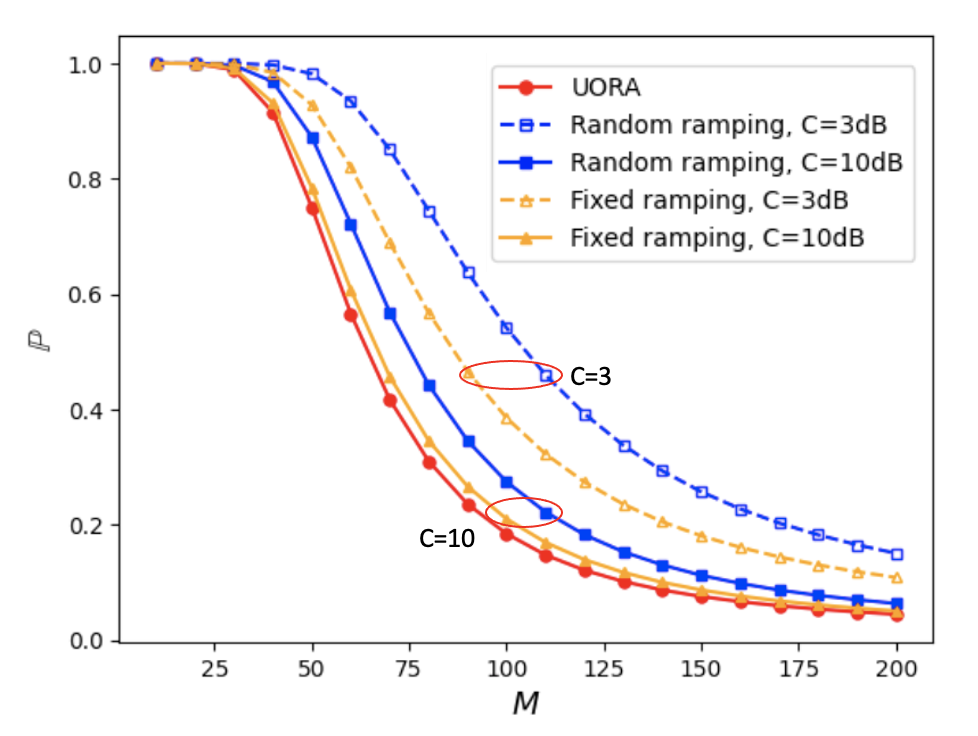}\label{fig:capture-p}}
\hfill
\subfigure[Average access delay]{\includegraphics[width=0.45\textwidth]{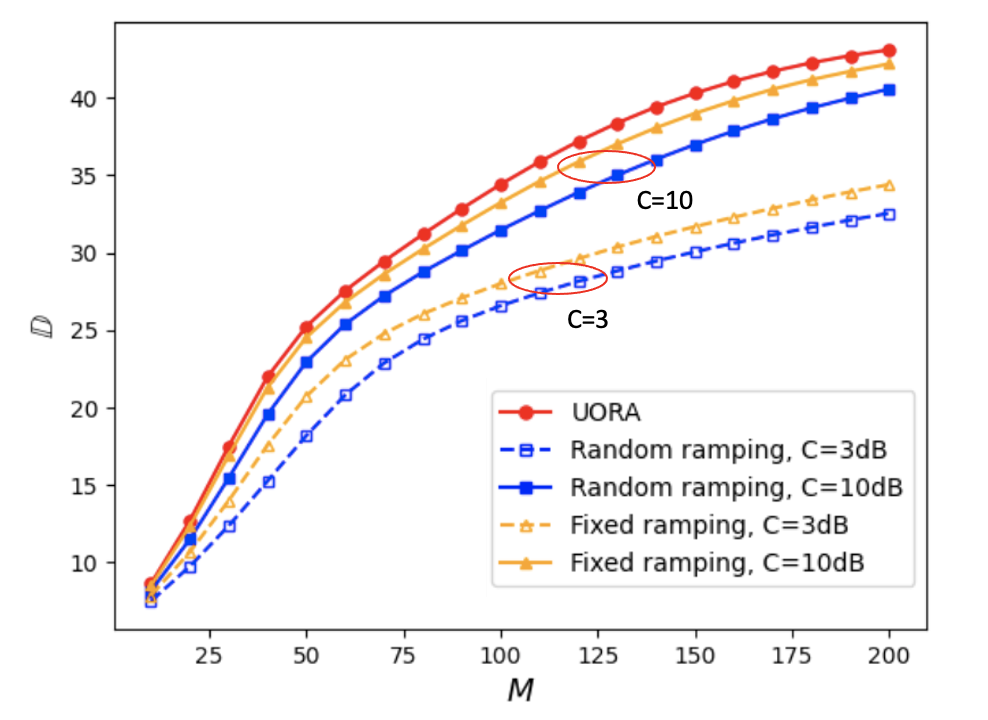}\label{fig:capture-d}}\\[-0.6ex]
\subfigure[RA-RU utilization]{\includegraphics[width=0.45\textwidth]{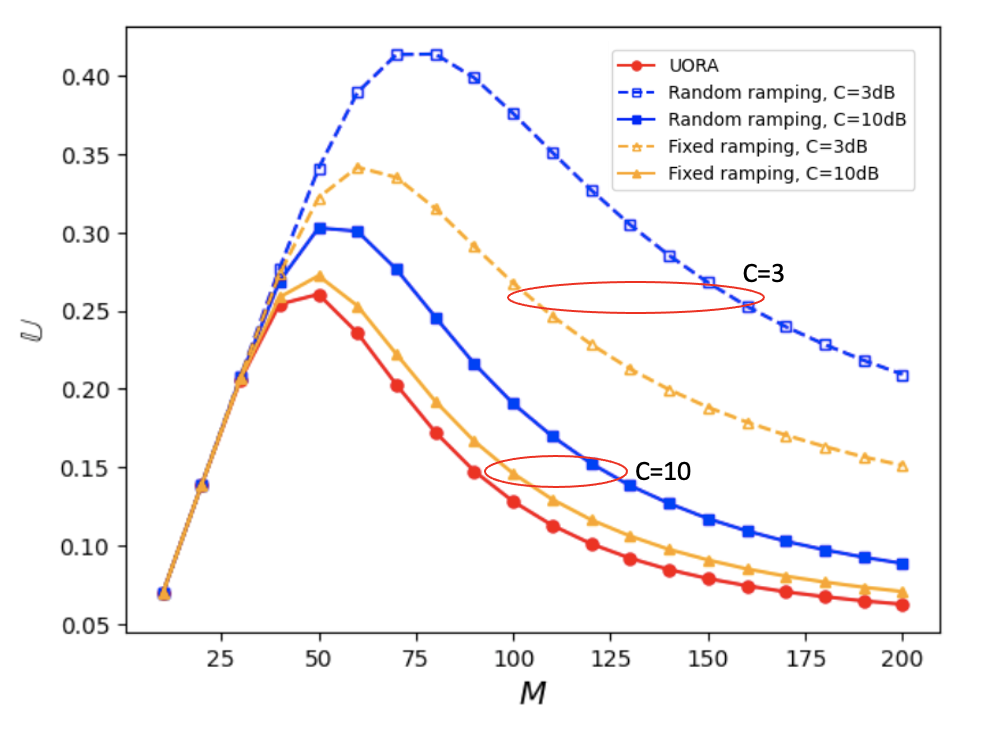}\label{fig:capture-u}}
\hfill
\subfigure[Power efficiency]{\includegraphics[width=0.45\textwidth]{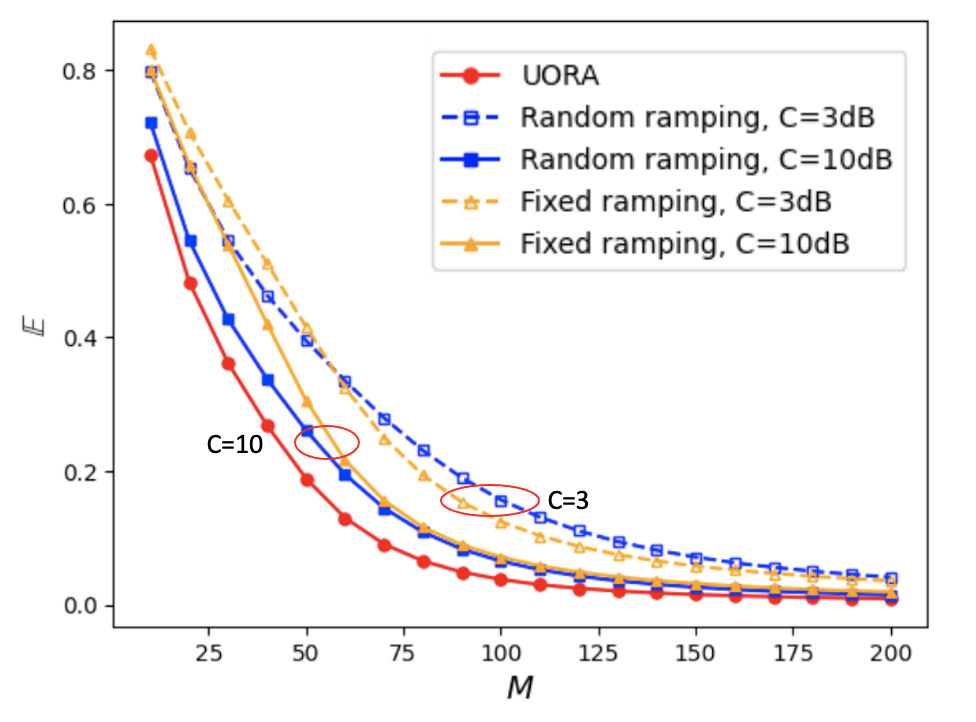}\label{fig:capture-e}}
\caption{Experiment: effect of the capture threshold on UORA and the two power-control schemes under $R=18$, $OCW_{\min}=7$, and $OCW_{\max}=31$. The four metrics are shown respectively.}
\label{fig:capture-threshold}
\end{figure*}

\section{Simulation Results}
Each data point is averaged over $10^5$ independent simulation samples. Unless otherwise stated, the baseline configuration in Table~\ref{tab:parameters} is used. The timing parameters are $T_{\textnormal{PIFS}}=0.025$ ms, $T_{\textnormal{SIFS}}=0.016$ ms, $T_{\textnormal{TF}}=0.1$ ms, $T_{\textnormal{RA-RU}}=5.484$ ms, and $T_{\textnormal{M-BA}}=0.032$ ms.

\begin{table}[!t]
\centering
\caption{Simulation parameters}
\scriptsize
\setlength{\tabcolsep}{2pt}
\renewcommand{\arraystretch}{1.06}
\begin{tabularx}{\columnwidth}{|c|X|c|}
\hline
Symbol & Description & Value \\
\hline
$C$ & Capture threshold; the strongest STA is decoded if its SIR exceeds $C$ & 3 dB \\
\hline
$R$ & Number of RA-RUs allocated by each trigger frame & 18 \\
\hline
$\gamma$ & Path-loss exponent in the received-power model & 2 \\
\hline
$OCW_{\min}$ & Initial OFDMA contention window size & 7 \\
\hline
$OCW_{\max}$ & Maximum OFDMA contention window size & 31 \\
\hline
$L_{\max}$ & Maximum number of transmission attempts and discrete power levels & 5 \\
\hline
$p_{\min}$ & Minimum STA transmit power & 10 mW \\
\hline
$p_{\max}$ & Maximum STA transmit power & 250 mW \\
\hline
$r$ & Radius of the circular service area & 10 m \\
\hline
\end{tabularx}
\label{tab:parameters}
\end{table}

Fig.~\ref{fig:capture-threshold} compares the effect of the capture threshold. When $C=3$ dB, both power-control schemes outperform fixed-power UORA after the network becomes moderately loaded. Random power level selection provides the highest success probability, the lowest delay, and the best RA-RU utilization because it increases the probability that one collided STA has sufficiently stronger received power. Step-wise power ramping is more power-efficient under light-to-moderate load because most retransmissions still occur at lower power levels. When $C$ increases to $10$ dB, capture becomes harder, so the gains of both power-control schemes move closer to conventional UORA.

%
%

\begin{figure}[H]
\centering
\subfigure[Average successful STA count across five equal-area rings]{\includegraphics[width=0.74\columnwidth]{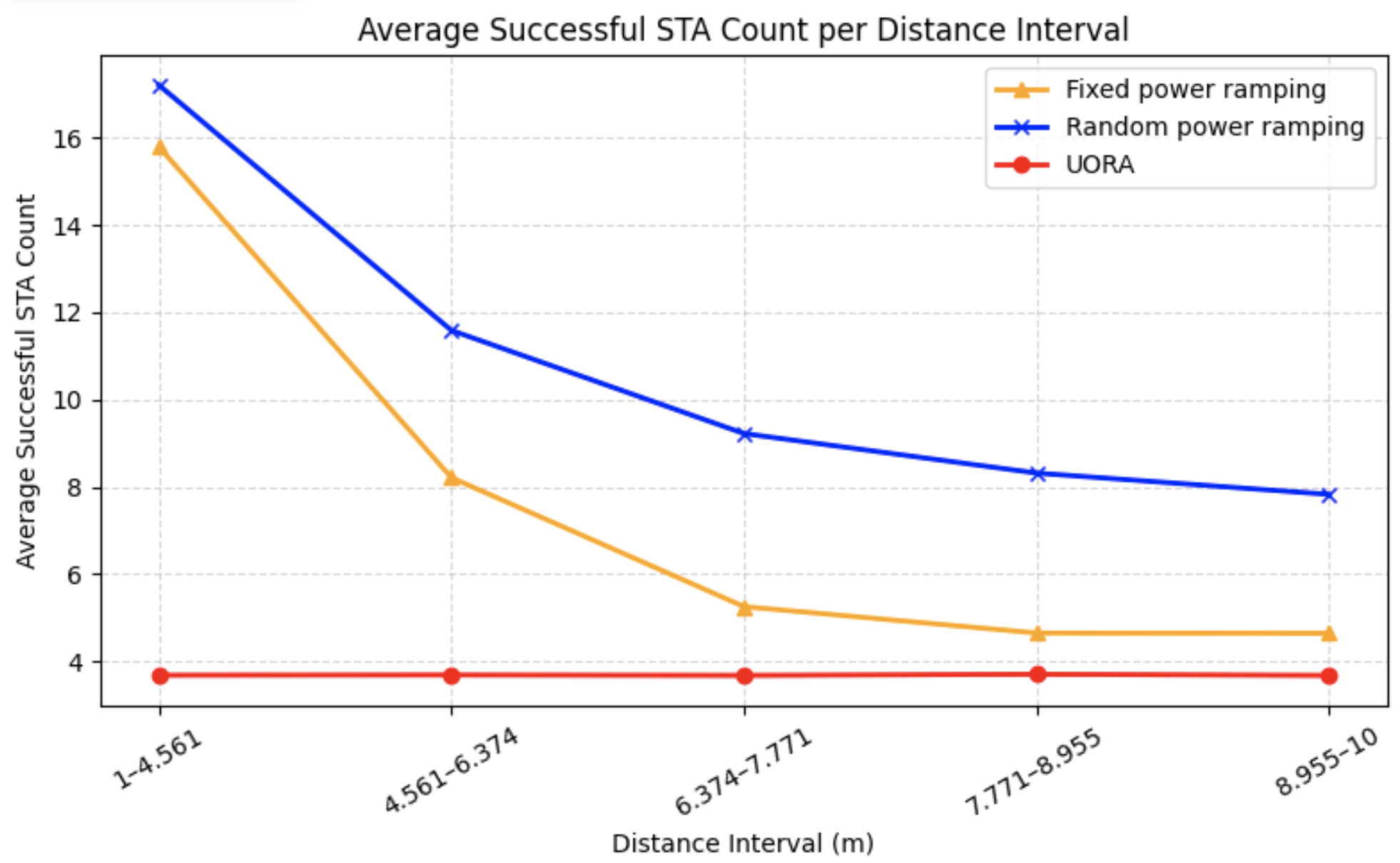}\label{fig:fair-count}}\\[-1.0ex]
\subfigure[Jain's Fairness Index]{\includegraphics[width=0.74\columnwidth]{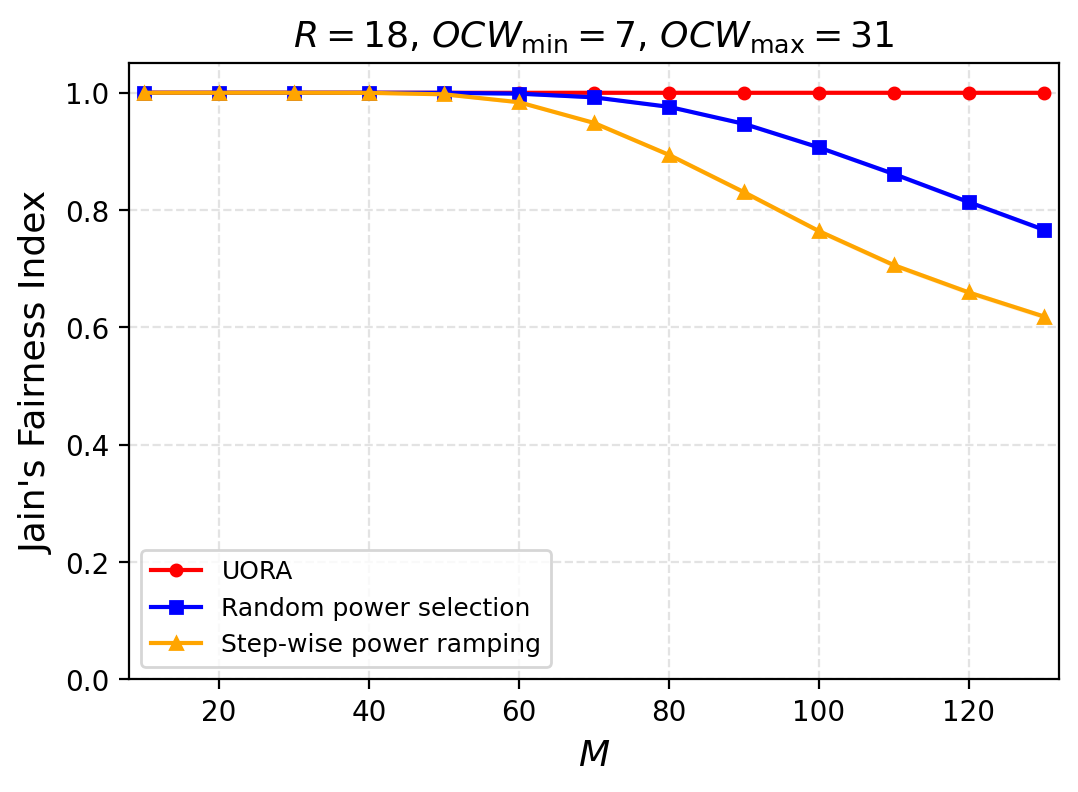}\label{fig:fair-jfi}}
\caption{Spatial fairness of capture-aware power control. Inner rings are closer to the AP, and outer rings are closer to the cell edge.}
\label{fig:fairness}
\end{figure}

Fig.~\ref{fig:fairness} shows the fairness impact of capture-aware power control. The service area is divided into five equal-area concentric rings from the AP to the cell edge. Under fixed-power UORA, the successful STA count is nearly uniform because, with $p_{\min}=10$ mW, $r=10$ m, and $\gamma=2$, even the cell-edge received power is above the receiver sensitivity. With capture-aware power control, inner-ring STAs obtain a persistent advantage because their received powers are larger during collisions. Jain's Fairness Index (JFI) is computed as
\begin{equation}
J=\frac{\left(\sum_{i=1}^{K}x_i\right)^2}{K\sum_{i=1}^{K}x_i^2},
\end{equation}
where $K=5$ and $x_i$ is the per-ring access success rate. Conventional UORA keeps $J\approx1$, while both power-control schemes reduce fairness under moderate-to-heavy load, especially step-wise power ramping.

\section{Conclusion}\label{sec:conclusion}
This paper evaluated random power level selection and step-wise power ramping for IEEE 802.11be UORA under the capture effect. Simulation results show that capture-aware power control can improve access success probability, reduce delay, and increase RA-RU utilization when the capture threshold is moderate. Random power level selection provides the strongest access-performance gain, while step-wise power ramping can provide better power efficiency under light-to-moderate load. The results also reveal a fairness tradeoff: capture-aware schemes favor STAs closer to the AP and reduce spatial fairness. Future work should therefore consider distance-aware or fairness-aware compensation mechanisms.

\section*{Acknowledgment}
This work was supported by the National Science and Technology Council (NSTC), Taiwan, under Contract numbers 112-2218-E-011-006 and 114-2221-E-011-078-MY3.


\begin{thebibliography}{00}

\bibitem{RTA applications}
E. Avdotin, D. Bankov, E. Khorov and A. Lyakhov, "Enabling Massive Real-Time Applications in IEEE 802.11be Networks," 2019 IEEE 30th Annual International Symposium on Personal, Indoor and Mobile Radio Communications (PIMRC), Istanbul, Turkey, 2019, pp. 1-6

\bibitem{adaptive backoff}
X. Zhu, Y. Zhong, W. He, L. Wang and T. Han, "Uplink OFDMA Random Access Mechanism Based on Adaptive Modulation Backoff Step," 2024 12th International Conference on Intelligent Computing and Wireless Optical Communications (ICWOC), Chongqing, China, 2024, pp. 113-117

\bibitem{Healthcare IoT}
S. Ould Amara and M. Yazid, "High Performance UORA Protocol Designed for IoT-Based Healthcare Applications Running on the EHT WLANs," 2024 1st International Conference on Innovative and Intelligent Information Technologies (IC3IT), Batna, Algeria, 2024, pp. 1-6

\bibitem{Double stage UORA}
S. O. Amara, M. Yazid and S. Mammeri, "Double-Stage UORA to Meet the Requirements of IOT Applications Operating Over Wi-Fi 7," 2024 4th International Conference on Embedded and Distributed Systems (EDiS), BECHAR, Algeria, 2024, pp. 28-33

\bibitem{Efficient Backoff}
K. Kosek-Szott and K. Domino, "An Efficient Backoff Procedure for IEEE 802.11ax Uplink OFDMA-Based Random Access," in IEEE Access, vol. 10, pp. 8855-8863, 2022

\bibitem{collision based}
A. Rehman, F. B. Hussain, R. Ali and T. Khurshaid, "Collision-Based Up-Link OFDMA Random Access Mechanism for Wi-Fi 6," in IEEE Access, vol. 11, pp. 117094-117109, 2023



\bibitem{bursty-11ax}
R. -G. Cheng, C. -M. Yang, B. S. Firmansyah and R. Harwahyu, "Uplink OFDMA-based random access mechanism with bursty arrivals for IEEE 802.11ax systems," in {\it IEEE Networking Letters}, vol. 4, no. 1, pp. 34-38, March 2022.

\bibitem{RL-UORA}
K. Kosek-Szott, S. Szott and F. Dressler, ``Improving IEEE 802.11ax UORA Performance: Comparison of Reinforcement Learning and Heuristic Approaches,'' in \textit{IEEE Access}, vol. 10, pp. 120285--120295, 2022.

\bibitem{UORA-optim}
P. Liu, E.-C. Park and J. Choi, ``Performance Optimization of IEEE 802.11ax UL OFDMA Random Access,'' in \textit{Journal of Communications and Networks}, vol. 26, no. 6, pp. 580--592, Dec. 2024.

\bibitem{MU-UORA}
J. Meng, Q. Zhao, W. Wu, M. Jin, P. Song and Y. Liu, ``Enhancing IEEE 802.11ax Network Performance: An Investigation and Modeling Into Multi-User Transmission,'' in \textit{IEEE Trans. Mobile Computing}, vol. 24, no. 3, pp. 2151--2165, Mar. 2025.

\bibitem{NOMA-RA}
J. Choi, ``NOMA-Based Random Access With Multichannel ALOHA,'' in \textit{IEEE J. Sel. Areas Commun.}, vol. 35, no. 12, pp. 2736--2743, Dec. 2017.

\bibitem{WiFi-ML-survey}
S. Szott, K. Kosek-Szott, P. Gaw{\l}owicz, J. Torres G\'{o}mez, B. Bellalta, A. Zubow and F. Dressler, ``Wi-Fi Meets ML: A Survey on Improving IEEE 802.11 Performance With Machine Learning,'' in \textit{IEEE Commun. Surveys Tuts.}, vol. 24, no. 3, pp. 1843--1893, 3rd Quart. 2022.












\end{thebibliography}
\end{document}